\documentclass[aps,pre,twocolumn,groupedaddress,showpacs,preprintnumbers, amsmath, amssymb]{revtex4}

\usepackage{graphicx}
\usepackage{bm}

\begin{document}

\title{Two-level system with a thermally fluctuating transfer matrix
element: Application to the problem of DNA charge transfer}

\author{Maria R. D'Orsogna}
\author{Joseph Rudnick}

\affiliation{Physics Department, University of California, Los Angeles, CA
90095-1547} 

\date{\today}

\begin{abstract}
Charge transfer along the base-pair stack in DNA is modeled in terms
of thermally-assisted tunneling between adjacent base pairs. Central to
our approach is the notion that
tunneling between fluctuating pairs 
is rate limited by the requirement of their optimal alignment.  
We focus on this aspect of the process by modeling two
adjacent base pairs in terms of a classical damped oscillator subject
to thermal fluctuations as described by a Fokker-Planck equation.  We
find that the process is characterized by two time scales, a result
that is in accord with experimental findings.
\end{abstract}

\pacs{87.15.-v, 73.50.-h,82.30.Fi}

\maketitle

\section{\label{sec:introduction}Introduction}

In spite of the fact that a decade has passed since the first
definitive observation of charge transfer along the DNA base-pair
stack \cite{murphy}, the detailed properties of this process have not
been definitively elucidated.  This is partly due to the inherent
complexity of the molecular structure of DNA, and to the large number
of external and intrinsic factors that exert an influence on DNA
structure and behavior.  The current unsettled situation also reflects
the absence of an overall agreement on the precise mechanism
by which this charge transport takes place.  One of the key issues
that awaits full illumination is the role of disorder---both static
and dynamic---on the propagation of charge along the base-pair
stack.  A related, and quite fundamental, question is whether charge
transport is a coherent quantum mechanical process, like conduction of
electronic charge against a static, or deformable, background, or
whether it it takes place as fundamentally incoherent transport, as a
variation of the random walk.  The answers to these and other
questions will have a significant impact on both our understanding of
the biological impact of charge transport in DNA and the development
of applications based on this phenomenon.

Despite the often contradictory results of experimental
investigations \cite{fink,porath,storm, pablo, giese}, 
a few conclusions seem inescapable.  The first is that
long-range charge transport along the base-pair stack depends quite
strongly on the sequence of the base pairs \cite{nunez}.  In addition,
base-pair mismatches can have a significant deleterious effect on
charge transport \cite{kelley,jackson} (see, however \cite{schuster}).
Furthermore, strands of DNA display considerable disorder, both static
\cite{calladinebook} and dynamic
\cite{brauns,swaminathan,cheat,troisi}.
Finally, several sets of experiments on 
ensembles of short DNA strands have
uncovered an unusual two-step charge transfer process 
\cite{barton99,zewail}.
These studies focus on fluorescent 
charge donors intercalated in DNA oligostrands.
As the charge migrates towards the acceptor,
the fluorescence is quenched
and the rate of migration is determined
by the decaying fluorescence profile.
The data reveals that this decay process 
occurs according to two characteristic
time-scales which 
are separated by more than an order of magnitude
\cite{barton99,zewail}.  Any model that purports to explain
charge transport must take all this into account.

In this paper, we discuss a model for short range 
charge transport along a base
pair stack that undergoes substantial structural fluctuations.  
The process
occurs via thermally-assisted 
quantum mechanical 
tunneling of charge carriers
from one base pair to the next,
under the assumption that
this tunneling is properly characterized 
as occurring in the presence
of a dissipative environment.  
A key conjecture is that charge transfer takes place
only when the neighboring pairs are in a state of
optimal ``alignment'',  and that
this alignment is statistically unlikely in thermodynamic equilibrium.
As we will see, this conjecture leads in a natural way to a model
exhibiting the dual-time-scale feature described above.  Additionally,
the model generates predictions that can be readily tested.
We shall relate the problem at hand
to the dynamics of a simple
two level system (TLS), realized by a donor and an acceptor state.

In Section~\ref{sec:tunneling}, we briefly recapitulate what is known
about the tunneling process in the presence of friction for a TLS system.
We also quantify our notion of a coordinate $\theta$
associated with the ``alignment'' of adjacent base pairs and of 
the influence of the dynamics of this new coordinate
on charge transfer. 
Section {\ref{sec:model}}
specifies the model for describing a generic collection of two-level
systems (TLS), initially in the donor state and characterized by a
fluctuating alignment variable $\theta$.  The probability distribution
of donor states, $W(\theta, \dot \theta, t)$ obeys a Kramers equation
with a sink term due to charge transfer to the acceptor. 
The rate of charge transfer will 
be expressed by the fluctuating rate $\Gamma(\theta)$.
This Kramers expression
is recast into the form of a Volterra equation with the use of a
Lie-Algebra approach defined on the Hilbert space of the eigenstates
of the Kramers equation for $\Gamma(\theta)=0$.  
We will discuss
limiting cases of the solution to obtain physical insight and to
reveal the two-time-scale decay of the probability distribution
due to the sink term. 
We conclude in
Section \ref{sec:conclusions} with a discussion of the possible
application of our results to charge transfer in strands of DNA
consisting of several base pairs.  

The key result of our calculations
lies in the determination of $P(\theta^*,t)$, 
the probability distribution of donor states
evaluated at the optimal configuration $\theta^*$ 
and with the $\dot \theta$ variable integrated out.
Indeed, under the assumption
that the tunneling 
process is most
effective at $\theta \sim \theta^*$,
this quantity is directly related to 
the fluorescence intensity $I(t)$
of the base pair complexes,
as probed by J. Barton and A. Zewail
\cite{barton99, zewail},
through the following:

\begin{equation}
\label{fluorescence}
I(t) = I_0 \left[1-\Gamma \int_{0}^{t} P(\theta^*,t') dt' \right].
\end{equation}

\noindent
The quantity $I_0$ of the above relationship is a 
proportionality constant
and $\Gamma$ is the integrated
rate of transfer to the acceptor.
We shall determine the double exponential 
character of $P(\theta^*,t)$, and hence of $I(t)$,
in qualitative agreement with
the experimental findings. 
The conjectures made on 
the existence of an optimal and unlikely
configuration $\theta^*$
will be crucial in obtaining the two stage decay process,
a result that justifies the 
assumptions made.

The model we shall construct is obviously
not restricted in application to DNA oligostrands.
Using our results, we may conclude that
in an ensemble of generic systems the migration
of a particle from donor to acceptor
proceeds statistically
as a two-time scale process, provided the transfering 
process is of rare occurrence.

\section{\label{sec:tunneling}The tunneling process}

The process of charge transfer from a donor site to an acceptor site
---a two-level system---is ubiquitous in biochemical and physical
phenomena \cite{datta}.  It occurs under a broad variety of
spatio-temporal conditions.  Chemical bond formation or destruction,
ATP production in photosynthetic reactions, or the operation of
semiconducting devices, all involve the transfering of charges to
and from specific sites, via thermal activation or quantum-mechanical
tunneling through an energy barrier.  Because of its intrinsic nature,
charge transfer via quantum-mechanical tunneling takes place on a
length scale of up to tens of angstroms \cite{muller}; larger
distances are possible if other transport mechanisms are involved.
These include thermal hopping among sites, which are typical in
disordered systems, the creation of conduction bands in metals, or of
lattice distortions of polaronic type in specific systems.

Quantum-mechanical tunneling from a donor site to an acceptor site is
quite simply represented by a two-level system (TLS) \cite{leggett}.
In this description, the tunneling particle is limited to being 
in the donor or in the acceptor state, while the other degrees of
freedom of the system, nuclear for instance, describe the charge
potential energy.  

\begin{figure}
\includegraphics[height = 2.5 in]{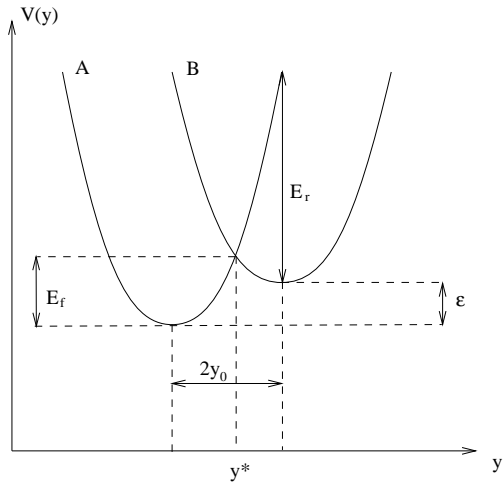}
\caption{This figure illustrates the nature of the tunneling
transition.  The two parabolic curves shown correspond to the two
versions of the potential $V( y, \sigma_{z})$ in Eq.
(\ref{spinboson}), one corresponding to the ``donor'' state in which
the tunneling particle is on one site and the other to the
``acceptor'' state, in which the particle is on the other one.
The two energies $E_f$ and $E_b = E_f-\epsilon$ 
referred to in the text are
the forward and backwards barrier energies respectively.
The horizontal axis corresponds to the reaction coordinate, $y$.}
\label{fig:flucdis}
\end{figure}

The energetic profile of the system
is thus characterized by a multidimensional surface of which
the acceptor and the donor states constitute relative minima,
separated by a barrier.
Of the many existing degrees of freedom, 
it is often possible to identify a ``reaction coordinate'' $y$
such that the energy barrier between donor and acceptor is
minimized along this specific direction. 
The progress of the reaction is then dominated by the
evolution along this coordinate and
the potential energy surface can be reduced to an
effective one-dimensional curve.

In certain systems the physical interpretation of the reaction
coordinate is immediate:  it may be the relative bond length in two
diatomic molecules, or solvent 
polarization around the donors and acceptors
\cite{barbara}. 
It is not an easy task to give a
physical interpretation of the reaction coordinate in the case of 
DNA base pairs because of 
the many possibilities involved - intra-base distance,
mobile counter-ion concentration, solvent concentration,
or a combination of all the above. A possibility is
offered by Ref. \cite{basko} where it is
suggested that the most relevant quantity 
is the interaction of the charge with the polar water molecules
of the solvent. 
In this paper we shall refer to the reaction coordinate $y$
in most general terms.

A common representation of tunneling with dissipation is through the 
spin-boson formalism \cite{leggett}.  The
donor and acceptor states are represented by means of a pseudo spin,
which points up when the charge is in the donor state 
and down otherwise.  
The Hamiltonian of the system is given by:

\begin{equation}
H_{ET}= \tau \sigma_x + \frac{P^2_y} {2M} + V(y, \sigma_z) + H_{\rm
bath},
\label{spinboson}
\end{equation}
\noindent
where 
\begin{equation}
V(y, \sigma_z) = \frac 1 2 M \omega ^2 (y+y_0 \sigma_z) ^2 +
\frac 1 2 \epsilon \sigma_z, 
\end{equation} 
and $\sigma_{x,z}$ are the Pauli
matrices.  The charge in the donor (up) state corresponds to the
potential $V(y, +)$ whose equilibrium reaction coordinate is $- y_0$,
and the converse state corresponds to $V(y,-)$, whose stable minimum
is at $y_0$.  The $H_{\rm bath}$ term represents contributions to
the Hamiltonian of a dissipative environment coupled to the reaction
coordinate.
Figure \ref{fig:flucdis} illustrates the meaning of the potential $V(
y, \sigma_{z})$ in the effective Hamiltonian of Eq.(\ref{spinboson}). The
curve marked A corresponds to the potential term in the donor state,
while the curve marked B represents the potential function in the
acceptor state.

This model has been thoroughly analyzed in the work by Garg {\textit et
al.} \cite{garg} based on earlier work by Leggett \cite{leggett2}.  A
similar analysis, but within a more chemical framework, is presented
by Marcus {\textit et al.} \cite{marcus-sutin}.
Energy conservation requires that charge
transfer takes place only when the reaction coordinate is close to the
degeneracy point $y=y^*$ for which $V(y^*,+) = V(y^*,-)$; once the
degeneracy point is reached, charge transfer is possible only
because of the non zero off-diagonal tunneling matrix elements $\tau$.

The tunneling rate $\Gamma$ from donor to
acceptor, is calculated in the above references.  For moderate
dissipation of the reaction coordinate, it is given by:

\begin{equation}
\label{gamma}
\Gamma = \frac{\tau^2}{\hbar}
\left( \frac{\pi}{E_r k_B T_{\rm eff}}\right)^{\frac 1 2} ~ 
\! \left( e^ {- E_f
/ k_B T_{\rm eff}} + e^{- E_b/ k_B T_{\rm eff}}\right)
\end{equation}

\noindent
where the reorganization energy $E_r$ and the energy barriers 
$E_f$ and $E_b$
depend on the details of the potential described by the reaction
coordinate.  
In the limit of high temperatures $T_{\rm eff}$ reduces
to the usual temperature $T$, whereas in the opposite limit the
quantity is temperature independent.  

The novelty explored in this paper is the introduction and
investigation of the effect of a second reaction coordinate, $\theta$,
governing the charge transfer process and coupled not to the energy,
but to the off-diagonal tunneling element $\tau$, hitherto been
treated as a constant, and which we now write $\tau(\theta)$.

This new
coordinate reflects the conjecture that in the case of DNA the
tunneling matrix element is highly sensitive to the donor-acceptor
relative configuration.
Charge transport along DNA in fact occurs along the
stacked base pairs by means of overlapping $\pi$
orbitals, and at room temperature, 
these base
pairs strongly fluctuate with respect to each other
through variations of the
twist, tilt and roll parameters \cite{calladine}.
The existence of base pair fluctuations for DNA in solution
is very well established, and is corroborated  
by experimental \cite{brauns} and molecular dynamics studies
\cite{swaminathan,cheat,troisi}.
For such a highly asymmetric system such as DNA,
fluctuations in the relative orientation
of donors and acceptors affect the magnitude of  
the orbital overlap between pairs, and
the new collective coordinate
$\theta$ embodies the effects of these fluctuations.

We will also assume that the
$\theta$ variable is slowly varying compared to the motion of the
reaction coordinate $y$, so as to define the lowest energy scale of
the system.  We may then separate the motion of the two reaction
coordinates in a Born-Oppenheimer spirit.
Charge transfer will be assumed to be instantaneous once
the optimal $\theta = \theta^*$ value is reached, and a
purely classical framework will be utilized for the $\theta$-dynamics.
The new reaction coordinate $\theta$ need not necessarily be pictured
as a geometrical one, although this is the framework we will be
utilizing in this paper.  Just as in the case of
the $y$ reaction coordinate,
$\theta$ may be associated to the
particular chemical environment of the molecule 
or to any other
quantity influencing the strength of the tunneling
element $\tau$ between the donor and the acceptor sites.

The Hamiltonian describing the system thus, is a modified version of
the spin-boson Hamiltonian introduced in Eq.  (\ref{spinboson}) with a
$\tau(\theta) \sigma_x$ off-diagonal term, as also described in
earlier work \cite{us}.  In order for charge transfer to take place,
we will assume that the reaction coordinate coupled to the energy must
be close to the degeneracy point $y=y^*$, and, also, that the $\theta$
coordinate must be in the neighborhood of an optimal value $\theta^*$,
which maximizes the tunneling amplitude.  The physical picture to
associate to this requirement is that the relative ``alignment'',
$\theta$, does not favor charge transfer unless an optimal
configuration is reached: $\tau(\theta) \simeq 0$ unless $\theta
\simeq \theta^*$.  
This conjecture will prove to be crucial in 
yielding the two time-scale charge transfer of references
\cite{barton99, zewail}.

In analogy to the experimental work cited above,
we consider a collection of such two-level systems, with the
charge initially located on the donor site.  
Each one of these
systems is associated to a particular $\tau(\theta)$ and through Eq.
(\ref{gamma}) to a particular $\Gamma (\theta)$ rate.
Our objective is to determine the mechanisms of charge transfer taking
into account the $\theta$ time evolution and the $\Gamma(\theta)$
rates accordingly distributed.  
We shall assume the $\theta$ dynamics to be governed
by small, Langevin type random fluctuations. 
At $t=0$, when the external charge is injected on the 
donor site, the distribution of $\theta$ values
is the usual Boltzmann distribution.
If the occurrence of the
optimal $\theta^*$ configuration
is relatively unlikely, we will indeed be able to 
show that the transfer process is characterized by a 
two time scale migration
of the initial donor population.

The emergence of two time scales in the transfer process
can be physically explained as follows.
The existence of an initial non-zero
population of TLS presenting the optimal value $\theta^*$, 
ensures that rapid tunneling to the acceptor.
The $\theta$ distribution is thus depleted of
population at the special value and other
transitions are forbidden to take place.  
The other TLS will tunnel to the acceptor
only after the system has re-equilibrated and re-populated
the optimal configuration, a process which is slow, because of the
assumption that the optimal configuration is a relatively unlikely
one. Hence, the existence of a fast, initial decay
followed by a slower decay process.

\section{The TLS and $\theta$ fluctuations}
\label{sec:model}

\subsection{The model}

Consider a collection of TLS which at the initial time $t=0$ are all
in the up-donor configuration, and characterized by the angular
parameter $\theta$.  Let us denote by $W(\theta, \dot \theta, t )$ the
TLS population remaining in the up-donor state at time $t$ and for
which the collective angular variable and its velocity are specified.

The physical requirement that $\theta$ be randomly, classically,
fluctuating in time, translates into the fact that $W(\theta, \dot
\theta, t )$ must evolve according to a Fokker-Planck type equation as
dictated by standard Langevin theory.  To this probability evolution
equation we must add an additional depleting term, that which
represents tunneling to the donor site as given by the
$\Gamma(\theta)$ term discussed above.

Different scenarios are possible for the $\theta$ dependence of $\tau$
and hence of $\Gamma$.  As discussed in the above section
we shall focus on the particular situation in
which tunneling is possible only for a very specific subset of
energetically unfavorable 
$\theta$ values.  In this picture,
tunneling is allowed only if donors and acceptors reach an
optimal---but unlikely---orientation one with respect to the other.
By including the tunneling term in the time evolution equation for
$W(\theta, \dot \theta, t)$ we obtain a modified Fokker-Planck
equation that may be used to approach any physical system in which the
presence of a depleting term competes with the usual Langevin
fluctuations.  The most natural choice for the $\theta$ motion, the
one we shall discuss in the remainder of this paper, is that of a
damped harmonic oscillator.  We shall see that starting from an
initially equilibrated system in which the $\theta$ distribution is the
the Boltzmann one, the insertion of the tunneling term will result in
the emergence of the two time scales discussed above.
We will refer to the
time derivative of the $\theta$ coordinate as $u$.  
The rotational moment of inertia associated
to $\theta$ is denoted by $I$ and its rotational frequency by
$\Omega$.

The goal of the next subsections will be to determine
$W(\theta^*, u, t)$, and in particular its integration
with respect to the $u$ variable. 
As described in the introduction in fact,
it is this quantity that is directly related
to the experiments we wish to model
by means of Eq.(\ref{fluorescence}).

\subsection{Kramers equation with a sink term}
\label{sec:FP}

\noindent The generic damped harmonic oscillator subject to random
noise responds to the following Langevin-type equations:
\begin{eqnarray}
\label{langevin1}
\dot \theta = u; \hspace{1cm} \dot u = -\gamma u - \Omega^2 \theta +
\eta(t),
\end{eqnarray}
\noindent where the stochastic force $\eta(t)$ is assumed to be a
zero-mean gaussian and whose correlation function is dictated by the
fluctuation-dissipation theorem for classical variables:
\begin{equation}
\label{flu-diss}
\left< \eta(t)\eta(t') \right> = \frac{2 \gamma k_B T}{I} \delta(t-t')
= 2q \delta(t-t').
\end{equation}
\noindent The corresponding Fokker-Planck equation may be written by
identifying \cite{risken} the proper co\-ef\-fi\-cients in the Kramers-Moyal
expansion from Eq.  (\ref{langevin1}) and is generally referred to as
the Kramers equation.  This equation governs the time evolution of the
distribution, $W(\theta, u, t)$, of an ensemble of systems obeying the
equations of motion (\ref{langevin1}).  It takes the form:
\begin{equation}
\label{kramers}
\frac{\partial W}{\partial t} = -u \frac{\partial W}{\partial \theta} +
\frac{\partial}{\partial u}[(\gamma u +\Omega^2 \theta) W] + q
\frac{\partial^2 W}{\partial u ^2}.
\end{equation}
\noindent 
The above equation is thoroughly analyzed
in \cite{Chandrasekhar}, where assuming an initial
probability distribution $W(\theta, u, 0) = \delta(\theta-\theta')
\delta(u-u')$, the probability $W(\theta,u,t)$, as well as other
relevant statistical quantities, are obtained.
At equilibrium Kramers equation is solved by the time
independent Boltzmann distribution,
$W(\theta,u,t)= \psi_{0,0}(\theta,u)$ with:
\begin{equation}
\label{ground1}
\psi_{0,0}(\theta,u) = \frac{\gamma \Omega} {2 \pi q} \exp
\left[-\frac{\gamma}{2q}\left(u^{2} + \Omega^{2}\theta^{2}\right)
\right].  \vspace{0.3cm}
\end{equation}
\noindent Under the assumptions discussed earlier, the probability
distribution function $W(\theta,u, t)$ for a particle localized on the
donor site and describing an effective angle $\theta$ with its
neighbor, will be described by the time evolution equation for a
collection of damped oscillators subject to a decay term $\Gamma$,
representing tunneling to the acceptor. The latter term is
appreciable only for a specific value of the $\theta$ coordinate
$\theta^*$:
\begin{equation}
\label{mine}
\frac{d W}{dt} = HW - \Gamma(\theta,u,t)~ W.
\end{equation}
\noindent The $H$ term is the differential operator that stems from
the right hand side of Eq.  (\ref{kramers}).  
We shall assume the decay term to be introduced at time $t=0$, prior
to which the system had attained its equilibration state.  In other
words, we choose the initial distribution $W(\theta,u,0)$ to be
Boltzmann-like, as expressed in Eq.  (\ref{ground1}).
For simplicity, we choose $\Gamma(\theta)$ to be independent
of $u$ and of $t$ and to be a gaussian centered
on $\theta^*$ and with width $\sigma$:
\begin{eqnarray}
\Gamma(\theta)=\frac{\kappa}{\sqrt{2\pi\sigma}}~ \exp
\left[-\frac{(\theta-\theta^*)^2} {2\sigma}\right].  \vspace{0.3cm}
\end{eqnarray}

\noindent 
The coefficient
$\kappa$ contains the physical parameters of
temperature and energy as expressed in Eq.  (\ref{gamma}).  We also
impose the constraint that at $t=0$ the optimal value $\theta^*$
carries a small Boltzmann weight.  This is equivalent to the physical
assumption that the occurrence of particle tunneling is a rather
unlikely event, and that the system tends to relax to $\theta$ values
that are far from the tunneling point.  We also impose the width of
the decay gaussian $\sqrt{\sigma}$, to be small compared to
$\theta^*$, so that $\Gamma(\theta)$ is highly peaked around the
optimal configuration value $\theta^*$: $\sqrt{\sigma} \ll \sqrt{q
/\gamma \Omega^2} \ll \theta^*$.

In the following subsections we will solve Eq.  (\ref{mine}) for
the early and long time regimes. The general solution for arbitrary times
is contained in the appendix.
The coupling of the system to the orientational degree of
freedom along the lines discussed above, manifests itself very clearly
in the unusual time dependence of the probability distribution.  Two
different decay rates in fact arise, with a rapid initial decay of the
donor population $W(\theta,u,t)$ followed by a slower transfer
process.  The ratio of these two time scales, and the main result of
this analysis is succinctly expressed by Eq.  (\ref{rates}) in terms
of all the physical parameters of this system.

\subsection{Short time regime}
\label{sec:shorttimes}

In order to determine the asymptotic behavior
of $W(\theta, u, t)$ in the early time regime, we
consider Eq. (\ref{mine}) with the gaussian choice of
$\Gamma(\theta)$  
and we perform a multiple time scale analysis
\cite{bender}. 
This is carried out by introducing a new ad-hoc 
variable $\xi = \Gamma(\theta) t$, 
into the probability distribution,
and by seeking solutions in the
form $W(\theta,u,t) = W_0(\theta,u,t,\xi) + \Gamma(\theta) ~
W_1(\theta,u,t,\xi) + \dots$. The 
Fokker-Planck equation is thus expanded 
in powers of $\Gamma(\theta)$
and, for the zeroth and first order terms, it yields:

\begin{eqnarray}
\label{homo2}
\frac{\partial W_0}{\partial t} - H W_0 &=&0, \\
\label{nonhomo2}
\frac{\partial W_1}{\partial t} - H W_1 &=& - \left[\frac{\partial
W_0}{\partial \xi} + W_0 \right] + u \Gamma^{-1} \frac{\partial
\Gamma}{\partial \theta} ~ W_1.
\end{eqnarray}

\noindent
Note that the partial derivative with respect to $t$ in the above
equations treats $\xi$ as an independent variable.  The solution to
the first equation is expanded in terms of the complete set
of functions
$\Psi_{m,n}(\theta,u,t)$ that solve
Eq. (\ref{homo2}) -  obtained in Eq. (\ref{general})
and Eq. (\ref{solutions}) of the appendix - with coefficients $A_{m,n}$ that
depend on $\xi$, i.e:

\begin{equation}
W_0(\theta,u,t) = \sum_{m,n} ~ A_{m,n}(\xi) ~ \psi_{m,n}(\theta,u) ~
e^{-\lambda_{m,n} t}.
\end{equation}

\noindent Substituting this solution for $W_{0}$ into Eq.
(\ref{nonhomo2}),  the inhomogeneous term in square brackets becomes:

\begin{equation}
-\sum_{m,n} ~ [\frac {\partial A_{m,n}}{\partial \xi} + A_{m,n}] ~
\psi_{m,n}(\theta,u) ~ e^{-\lambda_{m,n} t}.
\end{equation}

\noindent If this were the only term present on the right hand side of
Eq.  (\ref{nonhomo2}), then $W_1(\theta,u,t,\xi)$ would contain a
secular term in its solution of the type:

\begin{eqnarray}
\hspace{-0.2cm}
\label{solution2}
&& W_1(\theta, u, t) \sim \hspace{6cm} \\
\nonumber
&& \hspace{0.5cm}
-t \sum_{m,n} ~ [\frac {\partial A_{m,n}}{\partial \xi} + A_{m,n}] ~
\psi_{m,n}(\theta,u) ~ e^{-\lambda_{m,n}t}.
\end{eqnarray}

\noindent
Such a solution will eventually exceed the ``leading order'' one.
We determine the coefficients $A_{m,n}$ by requiring
that there be no secular term in the solution to the equation.
It is precisely this constraint that constitutes the underlying idea of
multiple scale analysis.  The above condition translates into
requiring that the non-homogeneous term within parenthesis in Eq.
($\ref{nonhomo2}$) or equivalently 
in Eq. (\ref{solution2}) vanish:

\begin{equation}
\frac{\partial A_{m,n}(\xi)}{\partial \xi} = - A_{m,n}(\xi).
\end{equation}

\noindent
We now solve for $A_{m,n}$. Im\-pos\-ing the 
ini\-tial con\-di\-tion $W(\theta,u,0) =
\psi_{0,0}(\theta,u)$ and reinserting $\xi = \Gamma(\theta) t$ the
solution reads:

\begin{equation}
\label{slow}
W_0(\theta,u,t) =\psi_{0,0}(\theta,u) ~ \exp ~ [-\Gamma(\theta) t].
\end{equation}

\noindent
The above is a zero-th order approximation
to the full problem presented in (\ref{nonhomo2}) to the extent that
the effect of $H$ acting on $t \Gamma (\theta)$ can be neglected with
respect to $\Gamma (\theta)$ itself.  In other words, 
Eq. (\ref{slow}) is an approximate solution as long as:

\begin{equation}
\label{limit}
t \ll \frac{\Gamma(\theta)}{|u \Gamma_\theta(\theta)|} =
\frac{\sigma}{|u (\theta-\theta^*)|}.
\end{equation}

\noindent This equation is valid only under the conditions expressed
in $(\ref{limit})$ and up to $t \simeq \Gamma^{-1}(\theta)$.
For this time limitation to be meaningful, it is necessary that the
width of the decay term $\sqrt{\sigma}$ be finite.  In the limit that
the width vanishes the above analysis fails, since the expansion parameter
diverges. At time $t \sim 0$ we cannot
approximate $\Gamma(\theta)$ by a strict delta function.
Note that for $\theta \sim \theta^*$, the tunneling point, and for
finite $u$ the condition arising from the multiple scale analysis ($t
\simeq \sqrt{2 \pi \sigma}/ \kappa$) is the most stringent one, and the
probability distribution is approximated by:

\begin{equation}
\label{short2}
W(\theta^*,u,t) = \psi_{0,0}(\theta^*,u) ~ \exp ~ [-\frac{\kappa
t}{\sqrt{2 \pi \sigma}}].
\end{equation}

\noindent We now perform an integration over the $u$
variable on both sides of Eq.  (\ref{slow}) and obtain an
approximation for the distribution probability function $P(\theta,t) =
\int_{-\infty}^{\infty} W(\theta,u,t) ~ du$:

\begin{equation}
\label{rate1}
P(\theta,t) \simeq \psi_{0}(\theta) ~ \exp ~ [- \Gamma(\theta)~ t].
\end{equation}

\noindent where $\psi_{0}(\theta)$ is the Boltzmann distribution
associated to the $\theta$ variable $\psi_{0}(\theta) =
\int_{-\infty}^{\infty} \psi_{0,0}(\theta,u) ~ du$.  For small times,
$P(\theta,t)$ retains its initial gaussian shape, with its amplitude
decreasing exponentially.

\subsection{Long time regime}
\label{sec:longtimes}

In this subsection we determine the long time 
asymptotic behavior of $W(\theta, u, t)$,
utilizing some of the results obtained in the appendix
for arbitrary times. In particular, we adapt the
kernel expansion of
Eq. (\ref{kernel0}) and Eq. (\ref{kernel})
to the long time regime.
Differentiating Eq.  (\ref{kernel0}) with respect to  $t$
and with the gaussian choice for $\Gamma(\theta)$ we obtain:
\begin{eqnarray}
\label{long}
\nonumber \frac {\partial W}{\partial t}= -\int \! \! \int
d \theta' du' \left.
\psi^{-1}_{0,0}(\theta',u') \frac{}{} \Gamma(\theta') \right.  \\
\nonumber \left[ \frac{}{} K(\theta,\theta',u,u',0) ~ W(\theta',u',t)
\right.  +\\
\left.  \int^{t}_{0} dt'~ \frac{\partial K}{\partial
t'}(\theta,\theta',u,u',t') ~ W(\theta',u',t-t') \right],
\\
\nonumber
\end{eqnarray}
\noindent 
where the integrals in $\theta'$ and in $u$ range from $-\infty$
to $+\infty$.
The time-derivative of the kernel in the last integral can
be obtained with the use of the expression obtained in Eq.
(\ref{kernel}) but with the summation restricted to non-zero values of
the integers $m$ and $n$.  
The contribution to the kernel of the term
associated with $m=n=0$ is time-independent,
and it has the form
$\psi_{0,0} (\theta,u) ~ \psi_{0,0} (\theta',u')$.  
We then replace
$\partial_{t'} K$ with $\partial_{t'} K'$ where $K'$ is defined as the
kernel without the first ($m,n=0$) summand.

The  function $K'$ and its time derivative contain exponentially
vanishing terms in $t$.  The time integrand in Eq.  (\ref{long}) will
therefore be appreciable only for $t' \leq \Omega_c^{-1}$ where
$\Omega_c$ is a cutoff frequency of the order of $|\lambda_{1,0}| =
\Omega$.  For $t \gg \Omega_c^{-1}$ we can approximate
$W(\theta',u',t-t') \simeq W(\theta',u',t)$ and restrict the time
interval from the origin to $\Omega_c^{-1}$.  Integrating by parts,
and using the above approximation for $W(\theta',u',t)$, the time
integral yields:
\begin{eqnarray}
\label{longmiddle0}
\frac {\partial W}{\partial t} &=& -\int \! \! \int
d \theta' du' ~
\psi^{-1}_{0,0}~ (\theta',u') ~ \Gamma(\theta') \\
&&
\nonumber
\left\{ \frac{}{} W(\theta', u', t)
\left[ \frac{}{}
K(\theta,\theta',u,u',0) + \right. \right. \\
&& 
\nonumber
 \left. \left. \frac{}{}
K'(\theta,\theta'u,u',\Omega_c^{-1})
-K'(\theta,\theta',u,u',0) \frac{}{} \right]\right\}.
\end{eqnarray}
\noindent
This equality is simplified by
$K'(\theta,\theta'u,u',\Omega_c^{-1})$ being negligible.
We can now rewrite the right hand side of
Eq. (\ref{longmiddle0}) as:
\begin{eqnarray}
\label{long2}
\frac {\partial W}{\partial t} = -\int \! \! \int d\theta' ~
du' \left[ \psi^{-1}_{0,0}(\theta',u')
\frac{}{} \Gamma(\theta') \right.\\
\nonumber \left.  \psi_{0,0}(\theta,u) ~ \psi_{0,0}(\theta',u')~
W(\theta',u',t) \frac{}{}\right].
\end{eqnarray}
\noindent
Since we are dealing with non-zero times, the
$\theta'$ integration can be performed under the assumption that
$\Gamma(\theta')$ is highly peaked around $\theta^*$ and
$\Gamma(\theta) \simeq \kappa ~ \delta(\theta-\theta^*)$:
\begin{eqnarray}
\frac{\partial{W}}{\partial{t}} =- \kappa ~ \psi_{0,0}(\theta,u)
\int_{-\infty}^{\infty} du' ~ W(\theta^*,u',t).
\end{eqnarray}
\noindent A last integration in the $u$ variable, performed on both
sides of the equation, yields the probability distribution function
for the $\theta$ variable:
\begin{equation}
\frac{\partial P(\theta,t)}{\partial t} = -\kappa ~ \psi_{0}(\theta) ~
P(\theta^*,t).
\end{equation}
\noindent For $\theta=\theta^*$ the above relationship yields a decay
rate of $-\kappa ~ \psi_{0}(\theta^*)$, and for arbitrary $\theta$
values we obtain the its behavior in the late time regime:
\begin{equation}
\label{rate2}
P(\theta,t) = P_0 ~ \psi_{0}(\theta) ~ \exp \left[- \kappa ~
\psi_{0}(\theta^*) ~ t \right].
\end{equation}

\noindent 
\subsection{The two time scales}

\noindent
As anticipated,
two different scenarios 
for $P(\theta^*, t)$ emerge from the analysis
carried out in the previous subsections.
From Eq. (\ref{rate1}), at early
times, the decay to the acceptor
state is rapid, occurring at a rate $r_1=\kappa /\sqrt{2 \pi
\sigma}$, whereas at latter times the rate is as given above: 
$r_2= \kappa \psi_{0}(\theta^*)$.  The ratio between the two is
\begin{equation}
\label{rates}
\frac{r_1}{r_2} = \sqrt{\frac{k_B T }{\sigma I \Omega^2 }} ~ \exp ~
\left[{\frac{I \Omega^2 }{2 k_B T } (\theta^*)^2} \right] \gg 1,
\end{equation}
\noindent as follows from the assumptions made on the gaussian
$\Gamma(\theta)$. The initial decay is much faster than that at
later times.

\subsection{Numerical results}
\label{sec:numbers}

Based on the general solution of Eq. ($\ref{kernel0}$),
we present a numerical analysis of 
the distribution function $W(\theta,u,t)$
for different choices of its arguments.  
In this equation the
probability distribution $W(\theta,u,t)$ is cast in a Volterra-type
formulation, for which solutions can be constructed iteratively in
time.  The probability distribution $W(\theta,u,t)$ as expressed in
Eq.  (\ref{kernel0}) in fact, depends only on its previous history and
on the known propagator function.

For a numerical approach, it is necessary to discretize the
$\theta,u,t$ variables and keep track of the value of $W(\theta,u,t)$
for every position and velocity at every temporal iteration.  While
feasible, this approach is rather cumbersome, since for every time
step $t_k = k \Delta t $ we must create a new $O(N^2)$ matrix
$W(\theta_i,u_j,t_k), ~ 1 \leq i,j \leq N$, where $N$ is the number of
spacings for the position and velocity meshes.  On the other hand, the
evaluation at of $W(\theta^*,u_j,t_k)$ where $\theta^*$ represents the
$\theta_i$ interval centered on the optimal value $\theta^*$ is
greatly simplified if the corresponding mesh is chosen so that
$\Gamma(\theta)$ may be replaced for all purposes by a delta function
at non-zero times.  The recursive equations now involve only the
$O(N)$ element vector $W(\theta^*, u_j, t_k), 1 \leq j \leq N$.
 
At $t=0$, when the propagator itself is a point source, the gaussian
shape for $\Gamma(\theta)$ must be retained for finiteness, but the
iteration at a time that is far from zero does not involve
values of the position that are significantly different from
$\theta^*$.  The $u$ mesh is chosen with $\Delta u = 0.05$ and the
time interval spacing is $\Delta t=0.01$.

\begin{figure}
\includegraphics[height=2.5in, angle=-90]{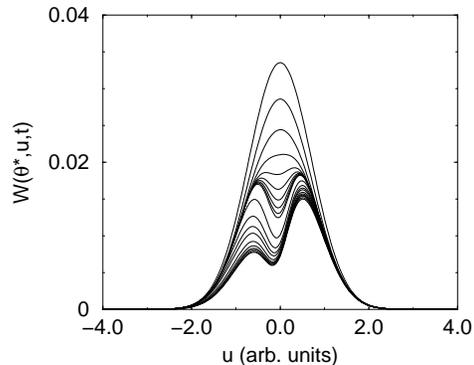}
\noindent 
\caption{The probability distribution at the optimal configuration 
$W(\theta^*,u,t)$ for various time intervals.
The top curve is evaluated at $t=0$ and is
the initial Boltzmann distribution, evaluated at
the unlikely configuration $\theta^*$.
The remaining
curves are its time evolution, up to
$t=5$ of the lower curve.}
\label{fig:dists}
\end{figure}

\noindent
In order to insure consistency with the constraint $\sqrt{\sigma} \ll
\sqrt{q/\gamma \Omega^2} \ll \theta^*$ we choose the following
parameters: $\sigma=10^{-4}$, $\gamma \Omega^2= 2q$, $\theta^*=1.5$.
The $\alpha$ parameter for the underdamped case is chosen as $\alpha =
0.02$, whereas $\kappa$ is fixed at $\kappa = 0.4$.  The resulting
probability distribution $W(\theta^*,u,t)$ is plotted in Figure
\ref{fig:dists} as a function of $u$ for various time intervals.

Two features of the evolving distribution are noteworthy.  The first
is the depression around $u=0$.  The second is a clear asymmetry in
the velocity distribution, in that the distribution for negative
values of the velocity, $u$, is lower than for positive $u$ values.
The reason for the first feature is the fact that when the velocity is
low, a pair will remain in a nearly optimal configuration longer, and
hence a tunneling event, leading to depletion of the distribution, is
more likely.  The asymmetry can be ascribed to the fact that the
optimal orientation is at positive values of the parameter $\theta$.
The time evolution equation encapsulates two mechanisms, one pushing
the distribution towards its Boltzmann limit, the other being the
tunneling process that leads to depletion of the distribution at
values of $\theta$ close to $\theta^{*}$.  In light of the trajectory
of the underdamped oscillation, a member of the ensemble with negative
velocity, $u$, is likely to be within a half an oscillation period of
having passed with a small velocity through $\theta^{*}$, which is
positive, while a representative with positive $u$ is more likely to
have spent more then half an oscillation period away from the optimal
tunneling configuration.  This latter, positive $u$ configuration will
have had more time to experience the ``restorative'' effects of the
mechanism that acts to generate the Boltzmann distribution.

\begin{figure}
\includegraphics[height = 2.5in, angle=-90]{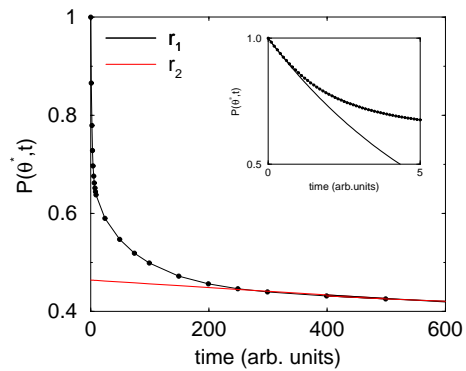}
\vspace{0.3cm} \caption{$P(\theta^*, t)$
for the parameters chosen in the text (dotted curves).  
The inset pertains to early times.   
The solid curves are drawn for comparison and
are exponential decays $e^{-rt}$, 
with a rate $r_1 = \kappa / \sqrt{2 \pi \sigma}$ 
for the short times of the inset, 
and $r_2 = \kappa \psi_0(\theta^*)$ for the long time regime.
Note the two distinct time scales.
}
\label{fig:pout2}
\end{figure}

It is also possible to perform a $u$-variable integration and obtain
the time dependence of $P(\theta^*,t)$.  The parameters are chosen as
above, and the two time scale decay of $P(\theta^*,t)$ can be clearly
seen to occur with rates $r_1$ and $r_2$ as described in Eq.
$(\ref{rates})$.  Also note that both at large and short times
$P(\theta^*,t)$ is proportional to $W(\theta^*,0, t)$.  The above
results, and the expressions for $r_1$ and $r_2$ are not affected by
changes in the damping variable $\alpha$.  
As anticipated, Figure $\ref{fig:pout2}$ clearly shows the double exponential
decay of $P(\theta^*, t)$, in agreement with 
the experimental results of \cite{barton99, zewail}.

\section{Discussion}
\label{sec:conclusions}

The model we have presented is expected to be of significant relevance
to charge transfer in DNA.
Thermal fluctuations 
strongly affect the structure of molecule,
and an accurate description 
requires this motion to be taken into account.

Not only has the existence of fluctuations
been experimentally documented \cite{brauns},
but it has also been suggested \cite{troisi}
that the motion that most affects the electronic coupling between
base pairs  - what we have referred to as $\tau(\theta)$ -
is their sliding one with respect to the other.
It must be pointed out that both these studies focus on
DNA in solution, not on dry strands of DNA. 

On the other hand, 
charge transport with more than one rate has been
reported in the literature  \cite{barton99}. 
For an oligomer
with the ethedium molecule acting as the donor, charge transfer
is found to occur along the same patterns as described by our model,
with two time scales of 5 and 75 picoseconds.
Two-time-scale decays are also observed in a series of
measurements \cite{zewail} performed on shorter strands of donor and
acceptor complexes (Ap-G). In these experiments 
the Ap donor can be treated, for
all practical purposes, as an intrinsic purine base,
and the ambiguity related to the choice of an extraneous
donor (the ethedium of the previous reference)
is removed.

In both these experiments,
an increase in the  length results in a 
competition between the fast and slow exponential decays in favor of
the slower time component.  
Increasing the length of the system 
diminishes the possibility that multiple base pairs
simultaneously arrange in the configuration
that facilitates rapid charge transfer.  When the process of
optimal alignment does occur (a relatively likely event only for a few
base pairs), the tunneling might not even require localization of the
charge on each base pair, and super-exchange can take place.

For long strands of DNA, thus, we expect 
the two intrinisic rates associated to a 
single charge transfer to be averaged out in favor of the slower
component.  
Traces of this unusual two time scale migration mechanism
however, may be found in the fact that DNA conductivity is enhanced
upon increasing the temperature \cite{gruner2}, presumably allowing
for greater base pair motion.  Charge transfer is also hindered by
disruptions to the stacking, which alter the base pair's ability to
find optimal transfer configurations, such as the insertion of bulges
along the helix or of strong mismatches within the base pair stacking
\cite{dandy,barton99_sci} which are poorly compatible with the
intrinsic conformation of the aromatic pairs.  Lastly, it is noted
that charge transfer effectiveness seems to be inversely proportional
to the measured hypochromicity \cite{barton97}, a quantity that
determines the ordering of base pairs along a certain direction and
defined as the reduction of absorption intensity due to interactions
between neighboring electric dipoles.  From this data it is apparent that
the higher the disorder of the system, the more efficient charge
transfer is.  
It would be interesting to see how different
solvent environments affect
conduction along the molecule in relation to their effect
on structural fluctuations. More temperature-dependent
experimental measures are desirable as well.

\section{Conclusions}
\label{sec:concludo}

We have presented a model for a spin boson TLS whose tunneling matrix
element depends on the structural conformation of the donor with
respect to the acceptor.  In the limit that the relative geometry
between the two fluctuates in time defining the lowest energy scale,
we are led to a classical problem, that of a collection of damped
harmonic oscillators obeying a modified Fokker-Planck equation.  If
charge transfer proceeds only for specific orientations of the donor
with  respect to the acceptors, the resulting rate for charge
transfer is divided into a fast component at short times and a
subsequent slower one.  These results agree with the experimental
findings of two-time-scale charge transfer in the donor intercalated
DNA complexes of J.Barton and coworkers \cite{barton99,zewail}.  It
must be noted that an implicit assumption of this work is that for
long range DNA conduction mediated by thermal fluctuations once the
charge has undergone a transfer between base pairs it does not return
to the pair at which it is originally localized.  However, it is
reasonable to assume that the transfer process will continue after
this event has occurred and that subsequent events will, with some
probability, deposit the charge at its point of origin at a later
time.  We have performed calculations on a two-time-scale hopping
model based on the results obtained here \cite{future}.  In these
calculations, the single set of two base pairs is replaced by a linear
array.  We have determined the probability that the charge carrier is
at its point of origin as a function of time, $t$, after its having
been placed there.  We find that this probability exhibits
two-time-scale behavior, with an initial, brief, rapid, exponential decay
followed by a much slower, power-law, decay at later times times.
The long-time asymptotics of this process are those of a random walk.

\begin{acknowledgments}
We acknowledge many useful conversations with Prof.  R. Bruinsma and
Prof.  T. Chou.
\end{acknowledgments}

\appendix
\section{General solution of the Kramers equation}
\label{sec:sol}

\noindent We shall adopt a Lie-Algebra approach \cite{note to Risken}
to identify a complete set of orthonormal functions that solve the
homogeneous problem in the general case of Eq.  (\ref{kramers}), and
through them the general solution for the decay equation (\ref{mine})
will be found.

Let us look for solutions of the following type, where $m$ and $n$
represent non negative integers:
\begin{equation}
\label{general}
\Psi_{m,n}(\theta,u,t)=\psi_{m,n}(\theta,u) e^{-\lambda_{m,n} t}.
\end{equation}
\noindent Upon in\-ser\-tion of the above
expression in Eq.  (\ref{kramers}) a time in\-de\-pen\-dent
Schr\"odinger-like equation can be written:
\begin{equation}
\label{homo}
-(\lambda_{m,n} + \gamma) \psi _{m,n} = H' \psi_{m,n},
\end{equation}
\noindent where:
\begin{equation}
H'(\theta, u) = q p_u^2 + \gamma u p_u + \Omega^2 \theta p_u - up_\theta,
\end{equation}
\noindent and the subscripts represent derivatives, $p_u = \partial
/\partial_u$.  As expected, the time independent Boltzmann
distribution satisfies the homogeneous equation, as can be verified by
direct substitution with $\lambda_{0,0}=0$.  The physical requirement
that solutions must be well behaved as $t \rightarrow \infty$, i.e.
that the $\lambda_{m,n}$'s be non negative, suggest that this is the
ground state:
\begin{equation}
\label{ground2}
\Psi_{\rm ground}(\theta,u,t) = \psi_{0,0}(\theta,u).
\end{equation}
\noindent The other solutions are found by constructing the ladder
operators.  For the underdamped case, we introduce the $\alpha$
variable such that $ \cos \alpha = \gamma/(2 \omega)$ and impose that
$[H', O]= l ~ O$ with $l$ and $O$ respectively complex variable and
operator to be determined.  
In practice, the operator $O$ corresponds
to either a raising or a lowering operator.  
Two sets of solutions
exist for the following `quanta' \ $l_{1,2}$:
\begin{equation}
l_1 = \Omega e^{-i \alpha}, ~ ~ l_2=\Omega e^{i \alpha},
\end{equation}
\noindent for which the associated raising and lowering operators
$R_{1,2}$ and $L_{1,2}$ are:
\begin{eqnarray}
\label{lo}
&& R_{1,2} = -p_\theta+ l_{1,2} ~ p_u; \\
&& L_{1,2} = \Omega^2 \theta +\frac{q}{\gamma} p_\theta +l_{1,2}
\left(\frac{q}{\gamma} p_u +u \right).
\end{eqnarray}
\noindent The commutation rules for the above operators can be easily
derived as:
\begin{eqnarray}
\nonumber && [R_i, R_j] =0, ~ [L_i, L_j]=0, ~ [R_1, L_2]=0, \\
&& {[ R_1, L_1]}= \Omega^2(e^{-2 i \alpha} -1), \\
\nonumber && {[ R_2, L_2]}= \Omega^2(e^{+2 i \alpha} -1).
\end{eqnarray}
\noindent The raising operators applied to the ground state yield the
set of solutions $\psi_{m,n}$ for Eq.  (\ref{homo}) with the
associated eigenvalues $\lambda_{m,n}$ as follows:
\begin{eqnarray}
\label{solutions}
\psi_{m,n} (\theta,u) = R_2^n ~ R_1^m ~ \psi_{0,0} (\theta,u), \\
\lambda_{m,n}= m \Omega e^{-i \alpha} + n \Omega e^{i \alpha}.
\end{eqnarray}
\noindent It is worth  noting that the Hamiltonian $H'$ can also be
reformulated as $ H'= (2 \Omega i \sin \alpha)^{-1} [(L_2 R_2)- (L_1
R_1)]$.
In order to construct solutions to the non-homogeneous problem within
the Hilbert space spanned by the set of solutions
$\{\psi_{m,n}(\theta, u)\}$, it is necessary to determine the
orthonormality of those solutions.  To this purpose, let us consider
the following $\{\phi'_{m,n}(\theta, u)\} = \{P_2^n P_1^m
\psi_{0,0}(\theta,u) \}$ where $P_{1,2}$ are operators defined as:
\begin{equation}
\label{pi}
P_{1,2} =-p_\theta-l_{1,2} ~ p_u.
\end{equation}
\noindent We can now prove an orthogonal relation between the two
sets, using the commutation rules and and introducing
$\psi_{0,0}^{-1}(\theta,u)$ as a weighting function:

\begin{eqnarray}
\nonumber
\int \int du ~ d\theta ~
\phi'_{m',n'}(\theta,u) ~ \psi_{0,0}^{-1}( \theta,u) ~
\psi_{m,n}(\theta,u) &=& \\ 
C_{m,n} ~ \delta_{m,m'} ~ \delta_{n,n'}.
\end{eqnarray}

\noindent
The integration limits are over the entire real axis, both for $\theta$
and $u$. The orthonormal set of eigenfunctions is thus expressed
as $\{ C_{m,n}^{-1} ~ \phi'_{m,n}(\theta,u) \}$, to which we refer as
$\{\phi_{m,n}(\theta,u) \}$. The constant of
proportionality $C_{m,n}$ is:
\begin{eqnarray}
C_{m,n} = m!  ~ n!  \left( \frac{\gamma \Omega^2}{q}\right) ^{m+n}
\left(1-e^{-2i\alpha}\right)^m
\left(1-e^{2i\alpha}\right)^n.
\end{eqnarray}
\noindent Let us now look for the full solution $W(\theta, u, t)$ 
to Eq.  (\ref{mine}), posing
it in the following form:
\begin{equation}
\label{nonhomo}
W(\theta,u,t)= \sum_{m,n} h_{m,n}(t) ~ \psi_{m,n} (\theta,u) ~
e^{-\lambda_{m,n} t }.
\end{equation}
\noindent The $h_{m,n}(t)$ functions are to be determined, in analogy
to the scattering problem of particles in quantum mechanics.  Let us
assume that the decay term is introduced at time $t=0$, and that the
initial distribution is the equilibrium solution to the homogeneous
problem, i.e. the ground state as expressed in Eq.  (\ref{ground1}).
Inserting Eq.  (\ref{nonhomo}) in Eq.  (\ref{mine}) and using the
orthonormality relations, it is possible to find time evolution
equations for $h_{m,n}(t)$ and to write a recursion formula for the
full solution:
\begin{eqnarray}
\nonumber W(\theta,u,t) &=& W(\theta,u,0) - \int_{0}^{t} dt'
\int_{-\infty}^{\infty} d\theta' \int_{-\infty}^{\infty} du' \\
\label{kernel0}
&& 
\nonumber
\left [\frac{}{}K(\theta,\theta',u,u', t-t') ~
\psi_{0,0}^{-1}(\theta',u') \right.  \\
&&  \left.\Gamma(\theta',u',t') ~W(\theta',u',t') \frac{}{}
\right].
\end{eqnarray}
\noindent 
Here, we have kept
$\Gamma$ a generic function of all variables
and the $K$ function is the response kernel of the
system:
\begin{eqnarray}
\nonumber 
\hspace{-1.5cm}
K(\theta,\theta',u,u', t) &=& \\
\label{kernel}
&& \hspace{-2cm}\sum_{m,n} \psi_{m,n}(\theta,u) ~ \phi_{m,n}(\theta',u')
~ e^{-\lambda_{m,n} t }.
\end{eqnarray}
\noindent The product $W'(\theta,u,t)= K(\theta,\theta',u,u',t) ~
\psi_{0,0}^{-1}(\theta',u')$, is the distribution function for the
homogeneous system, under the initial conditions $W'(\theta,u,0) =
\delta(\theta-\theta') \delta(u-u')$.  Its asymptotic behavior reduces
to the Boltzmann distribution, and apart from $t=0$, it is an
analytical function in all its variables.
The explicit representation of the kernel may be written by inserting
the expressions for {$\psi_{m,n}(\theta,u)$} and
{$\phi_{m,n}(\theta,u)$} in Eq.  (\ref{kernel}):

\begin{eqnarray}
\label{response1}
&& K\left(\theta,u, \theta^{\prime} u^{\prime},t \right) = \\
\nonumber \\
\nonumber && \exp \left[ \frac{q \left( \partial_\theta - \Omega e^{+i
\alpha}\partial_u \right) \left( \partial_{\theta^{\prime}} + \Omega
e^{+i \alpha}\partial_{u^{\prime}} \right)} {\gamma \Omega^{2} \left(
1 - e^{+2i \alpha}\right)} e^{- \Omega e^{+i \alpha}t} \right] \\
\nonumber \\
\nonumber && \exp \left[ \frac{q \left( \partial_\theta - \Omega e^{-i
\alpha}\partial_u \right) \left( \partial_{\theta^{\prime}} + \Omega
e^{-i \alpha}\partial_{u^{\prime}} \right)}{\gamma \Omega^{2} \left( 1
- e^{-2i \alpha}\right)} e^{- \Omega e^{-i\alpha}t} \right] \\
\nonumber \\
\nonumber && \hspace{3cm} \psi_{0,0}(\theta,u) ~
\psi_{0,0}(\theta'u'),
\end{eqnarray}

\noindent where the exponential terms are intended as operators acting
on the ground state wave functions.  As it is written, the above
kernel is still expressed symbolically. In order to obtain its
explicit form it will suffice to perform a Fourier transform of Eq.
(\ref{response1}), and then return to real space, a straightforward
but tedious calculation we omit.  The complete solution for the
kernel is given by \cite{note to Chandrasekhar}:
\begin{eqnarray}
&& K(\theta,\theta',u,u',t)= \left(\frac{\gamma \Omega}{2
\pi q}\right)^2 \frac{1}{\sqrt {T G}} \\
\nonumber \\
&& \nonumber \exp \left[-\frac{\gamma}{4 q T}
\left(\frac{}{}\Omega^2 (1-n)
(\theta+\theta')^2 + (1+l)(u-u')^2
\right.  \right.  \\
\nonumber &&\left.  \left.  \frac{}{} \hspace{3.5cm}
+2 m \Omega 
(\theta+\theta')(u-u') \right) \right] \\
\nonumber &&\exp \left[-\frac{\gamma}{4 q G}
\left(\frac{}{}\Omega^2 (1+n) (\theta-\theta')^2 \right.  \right.
\nonumber +(1-l) (u+u')^2 \\
&& \left.  \left.  \frac{}{}
\nonumber
\label{kernel3}
\hspace{3.5cm}+2m \Omega (\theta'-\theta)(u+u') \right) \right].
\end{eqnarray}

\noindent 
In order to keep a lighter notation,
we have suppressed
the time dependence of the $T(t)$, $G(t)$,
$l(t)$, $m(t)$, $n(t)$ functions.
They are defined as:
\begin{eqnarray}
l(t) \sin \alpha & = & e^{- \Omega t \cos \alpha} 
~ \sin(\alpha + \Omega t
\sin \alpha), \\
m(t) \sin \alpha & = & e^{-\Omega t \cos \alpha} 
~ \sin ( \Omega t \sin
\alpha), \\
n(t) \sin \alpha & = & e^{-\Omega t \cos \alpha} 
~ \sin (\alpha - \Omega t 
\sin \alpha).
\end{eqnarray}
The functions $T(t)$ and $G(t)$ are combinations  of the above:
\begin{eqnarray}
T(t) & = & 1+l(t)-n(t)-n(t)l(t) -m^2(t), \\
G(t) & = & 1+n(t)-l(t)-n(t)l(t)-m^2(t).
\end{eqnarray}

\noindent In order to ensure integrability for Eq.  (\ref{kernel0}),
some limitations are posed on the form of the $\Gamma(\theta',u',t')$
function.  For instance, the seemingly most natural choice, a delta
function centered around $\theta^*$, yields a non integrable
expression for $W(\theta,u,t)$ at small times, when the kernel is a
product of delta functions itself.  Instead, the gaussian choice
introduced earlier, with its finite $\sigma$, ensures integrability at
all time regimes.


\begin{references}
\bibitem{note to Risken}
    A slightly different method is also outlined in
    \cite{risken}, chapter 9.
\bibitem{note to Chandrasekhar}
    This solution, apart from the multiplicative
    $\psi_{0,0}^{-1}(\theta',u')$ term, is the same as discussed in
    \cite{Chandrasekhar}.  Integrating the $u$ coefficient, the
    expression is the same as Eq.(30) in Ref. \cite{Chandrasekhar}.
\end{references}
\end{document}